\documentclass[aps,prd,twocolumn,floatfix,superscriptaddress,longbibliography]{revtex4-2}
\usepackage{amsmath} 
\usepackage{graphicx,txfonts,bm}
\usepackage[colorlinks,citecolor=magenta,linkcolor=blue]{hyperref}

\begin{document}

\title{Reconstructing lattice QCD spectral functions with stochastic pole expansion and Nevanlinna analytic continuation}

\author{Li Huang}
\email{lihuang.dmft@gmail.com}
\affiliation{Science and Technology on Surface Physics and Chemistry Laboratory, P.O. Box 9-35, Jiangyou 621908, China}

\author{Shuang Liang}
\affiliation{Institute of Physics, Chinese Academy of Sciences, Beijing 100190, China}

\date{\today}

\begin{abstract}
The reconstruction of spectral functions from Euclidean correlation functions is a well-known, yet ill-posed inverse problem in the fields of many-body and high-energy physics. In this paper, we present a comprehensive investigation of two recently developed analytic continuation methods, namely stochastic pole expansion and Nevanlinna analytic continuation, for extracting spectral functions from mock lattice QCD data. We examine a range of Euclidean correlation functions generated by representative models, including the Breit-Wigner model, the Gaussian mixture model, the resonance-continuum model, and the bottomonium model. Our findings demonstrate that the stochastic pole expansion method, when combined with the constrained sampling algorithm and the self-adaptive sampling algorithm, successfully recovers the essential features of the spectral functions and exhibits excellent resilience to noise of input data. In contrast, the Nevanlinna analytic continuation method suffers from numerical instability, often resulting in the emergence of spurious peaks and significant oscillations in the high-energy regions of the spectral functions, even with the application of the Hardy basis function optimization algorithm.
\end{abstract}

\maketitle

\section{Introduction\label{sec:intro}}

Lattice QCD (LQCD) is a well-established first-principles and non-perturbative approach for studying strong interactions~\cite{lqcd_book,gupta1998,Wittig2020}. It serves as a valuable tool in understanding the genesis and evolution of the quark-gluon plasma (QGP)~\cite{annurev1996} and mapping out the phase diagram of strong-interaction matter~\cite{Ratti_2018,Lehner2022,AARTS2023104070}. In LQCD, spectral functions play a vital role in scrutinizing and elucidating high-energy physical phenomena that involve quarks and gluons, such as the melting of heavy quarkonium~\cite{PhysRevD.84.094504,PhysRevD.76.094513,PhysRevD.75.014506,PhysRevD.69.094507,PhysRevLett.92.012001,Umeda2005,Aarts2011,PhysRevD.86.014509} and the transport properties~\cite{PhysRevLett.99.022002,PhysRevD.83.034504,HE2023104020} of the QGP formed through relativistic heavy-ion collisions. However, accessing the spectral functions and other dynamical properties of the QCD medium from lattice simulations remains challenging due to LQCD's typical formulation on a discrete Euclidean space-time grid~\cite{gupta1998,lqcd_book,Wittig2020}. Therefore, researchers must reconstruct the spectral functions from numerically computed Euclidean correlation functions on the lattice to understand the relevant physics and compare the theoretical results with corresponding experimental data obtained from the Relativistic Heavy Ion Collider (RHIC) and the Large Hadron Collider (LHC)~\cite{TRIPOLT2019129}.

Mathematically, the Euclidean correlators $G(t)$ and the spectral functions $\rho(s)$ are connected through a Fredholm equation of the first kind: $G(t) = K(t, s) \circledast \rho(s)$, where $K(t, s)$ represents a continuous kernel function and $\circledast$ signifies convolution. Mapping the spectral functions to the Euclidean correlators is a straightforward process that can be easily accomplished using numerical integration. However, extracting spectral functions from Euclidean correlators through analytic continuation poses a formidable challenge~\cite{TRIPOLT2019129}. We observe that similar inverse problems are quite common in many-body and high-energy physics~\cite{many_body_book,lqcd_book}. They are considered ill-posed. There are two main reasons for this statement. Firstly, the Euclidean correlators are evaluated at a finite number of points due to the space-time discretization in LQCD~\cite{lqcd_book,gupta1998,Wittig2020}. Secondly, as LQCD simulations rely on stochastic Monte Carlo sampling, the resulting Euclidean correlators are inherently noisy~\cite{dipierro2001monte}. Even small deviations or fluctuations in the Euclidean correlators result in significant uncertainties in the spectral functions. As a result, the majority of resulting spectral functions exhibit high oscillations and lack physical significance, thereby inhibiting a reliable comparison between the theoretical spectra and experimental data. To address these issues, researchers have developed a plethora of analytic continuation approaches in the past decades. Next, we will briefly introduce some of the most commonly employed methods in LQCD simulations.

\emph{Maximum entropy method}. The maximum entropy method (MaxEnt) is perhaps the most popular analytic continuation tool, and it has dominated this field for a long time. In this method, the spectral density is interpreted as a probability distribution~\cite{JARRELL1996133,PhysRevB.44.6011,ASAKAWA2001459,10.3389/fphy.2022.1028995}. The primary objective is to extract the most probable spectral density $\rho$ from the correlation function $G$ and maximize the posterior probability $\text{Pr}[\rho|G]$. According to Bayes's theorem, $\text{Pr}[\rho|G] \propto \text{Pr}[G|\rho] \text{Pr}[\rho]$, where $\text{Pr}[G|\rho]$ is the likelihood function and $\text{Pr}[\rho]$ is the prior probability. It is important to incorporate analytical knowledge related to spectral properties in LQCD, such as positive definiteness or even the presence of pole structures within the spectra, into the probability distribution. A significant portion of the prior information could be encoded within the prior probability $\text{Pr}[\rho]$, which is proportional to $\exp{(\alpha S)}$, where $\alpha$ is a regulation parameter and $S$ denotes entropy. It is worth emphasizing that the entropic term $\alpha S$ is not unique and typically takes the form of the generalized Shannon-Jaynes entropy~\cite{JARRELL1996133,PhysRevB.44.6011}. Another popular alternative is the Bayesian reconstruction entropy~\cite{PhysRevLett.111.182003,10.3389/fphy.2022.1028995}. While the MaxEnt method is widely recognized for its efficiency and noise-tolerance, it sometimes struggles to faithfully recover sharp, subtle, and high-frequency features within the spectral functions.

\emph{Stochastic analytic continuation}. In the past decade, the stochastic analytic continuation method (SAC) and its variants have emerged as formidable contenders to surpass the MaxEnt method~\cite{PhysRevB.57.10287,beach2004,PhysRevB.62.6317,PhysRevE.94.063308,PhysRevB.76.035115,PhysRevE.81.056701,PhysRevX.7.041072,PhysRevB.101.085111,PhysRevB.102.035114,PhysRevB.78.174429,PhysRevD.97.094503}. Unlike the MaxEnt method, the SAC method treats all spectral functions equally instead of selecting the most probable one. Initially, the spectral functions are parameterized with hundreds or thousands of $\delta$-like functions. And then these parameters, such as the amplitudes and locations of the $\delta$ functions, are stochastically sampled at a fictitious temperature $\Theta$ using a Boltzmann-like weight function, which essentially serves as a likelihood function~\cite{PhysRevB.57.10287,beach2004}. Finally, the gathered spectral functions are filtered and averaged. We note that Shao and Sandvik \emph{et al.} have proven the equivalence in a generalized thermodynamic limit (large number of degrees of freedom) of the average spectrum and the maximum entropy solution~\cite{SHAO20231}. Although the SAC method supplements the MaxEnt method by enabling the resolution of subtle structures in spectra, it requires significant computational resources~\cite{PhysRevE.81.056701}.

\emph{Machine learning approaches}. In recent years, several machine learning aided methods have been developed to address the analytic continuation challenges in LQCD simulations. These methods include the deep neural networks (DNN)~\cite{PhysRevD.102.096001,SHI2023108547}, radial basis functions network (RBFN)~\cite{PhysRevD.104.076011}, entropy variational autoencoder (SVAE)~\cite{chen2022machine}, kernel ridge regression (KRR)~\cite{offler2021reconstruction}, automatic differentiation (AD)~\cite{PhysRevD.106.L051502}, Gaussian processes regression (GPR)~\cite{PhysRevD.105.036014}, and many others. Although these methods may offer improved performance in certain situations, their universality is not guaranteed. Furthermore, some studies have adopted supervised approaches to train machine learning network models, which incorporate prior knowledge from specific physics insights into the training sets. However, caution must be exercised as there is a risk of introducing biases in the training data.

Recently, two new analytic continuation methods, namely the stochastic pole expansion (SPX)~\cite{huang2023stochastic} and the Nevanlinna analytic continuation (NAC)~\cite{PhysRevLett.126.056402,PhysRevB.104.165111}, have been proposed. The SPX method inherits the spirit of the SAC method, where the Matsubara Green's function is initially parameterized with hundreds or thousands of poles. Subsequently, the amplitudes and positions of these poles are optimized using a stochastic algorithm that based on stimulated annealing~\cite{SA1983}. The SPX method is applicable to fermionic and bosonic systems. It has been extended to support analytic continuation of matrix-valued Green's functions~\cite{huang2023stochastic}. On the other hand, similar to the Pad\'{e} approximation (PA)~\cite{Vidberg1977,PhysRevB.93.075104,PhysRevB.61.5147,PhysRevB.87.245135}, the NAC method aims to interpolate the Matsubara data in the complex plane using some form of continued fraction expansion~\cite{PhysRevLett.126.056402,PhysRevB.104.165111}. It takes the ``Nevanlinna'' analytic structure of the Matsubara Green's function into consideration, ensuring that the calculated spectral functions are inherently positive and normalized. However, this method is highly sensitive to the noise level in the raw Matsubara data. With noiseless data as input, it can successfully resolve complex spectral functions across a wide energy range with unprecedented accuracy. Unfortunately, when the Matsubara data contains noise, the Nevanlinna interpolants may not exist, and the resulting spectral functions are not guaranteed to be causal.

Both the SPX and NAC methods have not yet been employed in addressing the issue of analytic continuation in the LQCD simulations, and it remains uncertain whether they are applicable for the analytic continuation of Euclidean correlation functions~\cite{TRIPOLT2019129}. Therefore, the purpose of this study is to fill this knowledge gap and expand the potential applications of these methods. We first generate noisy Euclidean data using four representative models: the Breit-Wigner model, the Gaussian mixture model, the resonance-continuum model, and the bottomonium model. These synthetic data sets are then processed using the SPX, NAC, and MaxEnt methods. Finally, we conduct a comprehensive comparison between the calculated spectra and the exact solutions if available. The results suggest that the SPX method manifests comparable or even superior performance in comparison to the commonly used MaxEnt method. On the other hand, the NAC method tends to suffer from numerical instability, even in the absence of noise in the input data.

The structure of the remaining sections of this paper is as follows: Section~\ref{sec:method} provides an introduction to the basic formulations of Euclidean correlation functions and offers a brief overview of the SPX and NAC methods. In Section~\ref{sec:test}, we present the computational setups, demonstrate, and discuss four representative examples. Section~\ref{sec:dis} explores the robustness of the SPX method and numerical instability of the NAC method in the presence of noisy Euclidean data. Additionally, we analyze the effects of the constrained sampling algorithm and the self-adaptive sampling algorithm on the SPX method, along with the impact of the Hardy basis function optimization algorithm on the NAC method. Finally, our findings are summarized in Section~\ref{sec:con}.

\section{Method\label{sec:method}}

\subsection{Euclidean correlation function}

At finite temperature, the Euclidean correlation function $G(\tau)$ is related to the spectral function $\rho(\omega)$ through~\cite{TRIPOLT2019129}
\begin{equation}
G(\tau) = \int^{\infty}_0 \text{d}\omega~\rho(\omega) 
    \frac{\cosh{[\omega(\tau-\beta/2)]}}{\sinh{(\beta\omega/2)}}.
\end{equation}
Here $\beta = 1/T$ represents the inverse system temperature, and $\tau$ represents the Euclidean (imaginary) time interval ($\tau \in [0,\beta]$). In momentum space, the expression for the Euclidean correlation function is~\cite{gelis_2019}
\begin{equation}
\label{eq:kl}
G(p) = \int^{\infty}_0 \text{d}\omega~\rho(\omega)
    \frac{\omega}{\omega^2 + p^2},
\end{equation}
where $p$ represents the Euclidean (Matsubara) frequency, the value of $G(p)$ can be derived by performing a discrete Fourier transform on $G(\tau)$. In the references, Eq.~(\ref{eq:kl}) is also known as the K$\ddot{\text{a}}$ll$\acute{\text{e}}$n-Lehmann (KL) spectral representation~\cite{qft_book}. By applying the process of analytic continuation, the retarded propagator $G^{R}(\omega)$ can be attained, enabling the extraction of the spectral function through the following expression:
\begin{equation}
\label{eq:rho}
\rho(\omega) = -\frac{1}{\pi} \text{Im} G^{R}(\omega),
\end{equation}
where $\omega = -ip$. It is important to note that $\rho(\omega)$ must be an odd function for bosonic system, i.e., $\rho(-\omega) = -\rho(\omega)$.

\subsection{Stochastic pole expansion}

According to the textbooks of many-body physics, the Lehmann representation of the finite temperature many-body Green's functions is given by the following formula~\cite{many_body_book,many_body_book_2016}:
\begin{equation}
\label{eq:lehmann}
G(z) = \frac{1}{Z} \sum_{m,n} \frac{\langle n | d | m \rangle \langle m | d^{\dagger} | n \rangle}{z + E_n - E_m} \left(e^{-\beta E_n} \pm e^{-\beta E_m}\right).
\end{equation}
In this expression, $d$ and $d^{\dagger}$ represent the annihilation and creation operators, respectively. $|n\rangle$ and $|m\rangle$ are the eigenstates of the Hamiltonian $\hat{H}$, and $E_n$ and $E_m$ are the corresponding eigenvalues. $Z = \sum_n \exp(-\beta E_n)$ is the partition function. $z \in \mathbb{C} \backslash \mathbb{R}$. The positive sign corresponds to fermions, while the negative sign corresponds to bosons. By introducing $A_{mn} = \langle n | d | m \rangle \langle m | d^{\dagger} | n \rangle \left(e^{-\beta E_n} \pm e^{-\beta E_m} \right) / Z $ and $P_{mn} = E_m - E_n$, Eq.~(\ref{eq:lehmann}) can be simplified as:
\begin{equation}
G(z) = \sum_{m,n} \frac{A_{mn}}{z - P_{mn}}.
\end{equation}
It is evident that only terms where $A_{mn} \neq 0$ can be present. The indices $m$ and $n$ can also be compressed as $\gamma$, resulting in the following expression:
\begin{equation}
\label{eq:pole}
G(z) = \sum^{N_p}_{\gamma = 1} \frac{A_{\gamma}}{z - P_{\gamma}}.
\end{equation}
Eq.~(\ref{eq:pole}) is referred to as the \emph{pole representation} of the many-body Green's functions~\cite{many_body_book}. In this representation, $N_p$ denotes the number of poles, and $A_{\gamma}$ and $P_{\gamma}$ mean the amplitude and position of the $\gamma$-th pole. For the Euclidean correlation in momentum space, its pole representation can be reformulated as:
\begin{equation}
\label{eq:pole_qcd}
G(p) = \sum^{N_p}_{\gamma = 1} \Xi(p, P_{\gamma}) \tilde{A}_{\gamma}.
\end{equation}
Here, $\Xi$ represents the kernel matrix, which is calculated using the following equation:
\begin{equation}
\Xi(p,\omega) = -\frac{G(0)\omega}{p - \omega}.
\end{equation}
$\tilde{A}_{\gamma}$ is the renormalized amplitude of the $\gamma$-th pole, given by:
\begin{equation}
\tilde{A}_{\gamma} = -\frac{A_{\gamma}}{G(0)P_{\gamma}}.
\end{equation}
It can be easily proven that $\tilde{A}_{\gamma}$ and $P_{\gamma}$ must satisfy the following constraints:
\begin{equation}
\label{eq:constraints}
\forall \gamma,~0 \le \tilde{A}_{\gamma} \le 1,~\sum_{\gamma} \tilde{A}_{\gamma} = 1,~\text{and}~P_{\gamma} \in \mathbb{R}.
\end{equation}

We assume that the input Euclidean correlation function is denoted as $\mathcal{G}(p_n)$, and the input data consists of $N$ frequency points. We then utilize Eq.~(\ref{eq:pole_qcd}) to approximate the Euclidean data. To assess the discrepancy between $\mathcal{G}(p_n)$ and $G(p_n)$, we introduce the so-called goodness-of-fit function $\chi^{2}$. Its definition is as follows: 
\begin{equation}
\label{eq:loss}
\chi^{2}
\left[\left\{\tilde{A}_{\gamma}, P_{\gamma}\right\}^{N_p}_{\gamma = 1}\right]
= \frac{1}{N}\sum^{N}_{n = 1}
\left|\left|
\mathcal{G}(p_n) - \sum^{N_p}_{\gamma = 1} \Xi(p_n,P_{\gamma}) \tilde{A}_{\gamma}
\right|\right|^2_{F},
\end{equation}
where $|| \cdot ||_{F}$ represents the Frobenius norm. Hence, the objective of the analytic continuation is to solve the subsequent multivariate optimization problem:
\begin{equation}
\label{eq:chi2}
\mathop{\arg\min}\limits_{ \left\{\tilde{A}_{\gamma}, P_{\gamma}\right\}^{N_p}_{\gamma = 1} } \chi^{2}\left[\left\{\tilde{A}_{\gamma}, P_{\gamma}\right\}^{N_p}_{\gamma = 1}\right].
\end{equation}
Once the optimized parameters $N_p$, $\tilde{A}_{\gamma}$, and $P_{\gamma}$ are determined, evaluating the retarded Green's function is straightforward by substituting $p$ with $\omega + i0^{+}$ in Eq.~(\ref{eq:pole_qcd}). Additionally, the spectral function $\rho(\omega)$ is computed using Eq.~(\ref{eq:rho}). It is important to note that this optimization problem [i.e., Eq.~(\ref{eq:chi2})] is highly non-convex. Traditional gradient-based optimization methods typically fail to identify the global minimum unless the initial solution is of high quality~\cite{optimization_book}. Therefore, in the SPX method, we employ the simulated annealing algorithm~\cite{SA1983} to optimize the $\tilde{A}_{\gamma}$ and $P_{\gamma}$ parameters subject to the constraints defined by Eq.~(\ref{eq:constraints}). For technical details regarding the possible Monte Carlo random walking rules in the configuration space $\mathcal{C} = \{\tilde{A}_{\gamma}, P_{\gamma}\}$, please refer to Ref.~\cite{huang2023stochastic}. The advantages of the SPX method include its ability to derive approximate expressions for correlation functions and its ease of extension to support the analytic continuation of bosonic systems, two-particle Green's functions, matrix-valued Green's functions, and so on. Application to noisy Matsubara data suggests that the SPX method can accurately resolve both continuum spectra for condensed matter cases and multiple $\delta$-like peaks for molecule cases. Notably, it performs well in reproducing sharp high-frequency features.

\subsection{Nevanlinna analytic continuation}

It is well known that the retarded Green's function, denoted as $G^{R}(\omega + i0^{+})$, and the Matsubara Green's function, denoted as $G(i\omega_n)$, can both be consistently represented as $G(z)$, where $z \in \mathbb{C} \backslash \mathbb{R}$. The NAC method utilizes the fact that the negative fermionic Green's function, denoted as $f(z) = -G(z)$, belongs to the class of Nevanlinna functions. By applying the invertible M$\ddot{\text{o}}$bius transform $h(z) = (z-i)/(z+i)$ to the function value of $f(z)$, the Nevanlinna function is mapped in a one-to-one fashion to a contractive function $\theta(z) = h[f(z)]$. This contractive function $\theta(z)$ can be expressed in the form of a continued fraction expansion, and an iterative algorithm can be constructed accordingly~\cite{PhysRevLett.126.056402}. The recursion relation between two steps $\theta_j(z)$ and $\theta_{j+1}(z)$ is given by:
\begin{equation}
\label{eq:recursion-relation-theta}
\theta_j(z) = \frac{ \theta_{j+1}(z) + \gamma_j}{\gamma_j^*h_j(z) \theta_{j+1}(z) +1}.
\end{equation}
In this equation, $h_j(z) = (z-Y_j)/(z+Y_j)$, $Y_j = i\omega_j$ represents the $j$-th Matsubara frequency used, and $\gamma_j = \theta_j(Y_j)$ represents the function value of the $j$-th contractive function at the point $Y_j$. The final expression of the recursive function $\theta(z)$ can be written as~\cite{10.1155/S1687120003212028}:
\begin{equation}
\label{eq:recursive-theta}
\theta(z)[z;\theta_{N_s+1}(z)] = \frac{a(z)\theta_{N_s+1}(z) + b(z)}{c(z)\theta_{N_s+1}(z) + d(z)},
\end{equation}
where
\begin{equation}\label{eq:factor-matrix}
  \left(
    \begin{matrix}
      a(z) & b(z) \\
      c(z) & d(z)
    \end{matrix}
  \right) = \prod_{j=1}^{N_s}
  \left(
    \begin{matrix}
      h_j(z)           & \gamma_j \\
      \gamma_j^*h_j(z) & 1
    \end{matrix}
  \right),
\end{equation}
with $j$ increasing from left to right. Here $N_s$ is the overall iteration step, which is equivalent to the number of data points. After obtaining $\theta(z)$, one can immediately get the Green's function by an inverse $\rm{M\ddot{o}bius}$ transform as $G(z)= -h^{-1}[\theta(z)]$. Note that the Pick criterion~\cite{Pick1917} should be fulfilled for the existence of the Nevanlinna interpolation.

Additionally, it is worth noting that there is flexibility in choosing $\theta_{N_s+1}(z)$, which can be used to select the most desirable spectral function. In Reference~\cite{PhysRevLett.126.056402}, $\theta_{N_s+1}(z)$ is expanded in the Hardy basis and chosen in such a way that it achieves the smoothest possible spectral function~\cite{Krantz1999}. The loss function employed in this selection process is given by:
\begin{equation}
\label{eq:loss-function}
\mathcal{L} = \left\vert 1-\int \frac{{\rm{d}} \omega}{2\pi} \rho_{\theta_{N_s+1}}(\omega) \right\vert^2 + \lambda \left\Vert \frac{{\rm{d}}^2 \rho_{\theta_{N_s+1}}(\omega)}{{\rm{d}} \omega^2} \right\Vert^2_F.
\end{equation}
This loss function consists of two terms. The first term enforces the proper sum rule, while the second term incorporates the smoothness condition. $\lambda$ is an adjustable parameter. By preserving the ``Nevanlinna'' analytic structure of Green's functions, the NAC method automatically generates positive and normalized spectral functions~\cite{PhysRevLett.126.056402}. However, it is important to emphasize that the method is sensitive to noise, and either a large number of data points $N$ or a high Hardy order $H$ can potentially lead to numerical instabilities.

Although the NAC method has been extended to support the analytic continuation of matrix-valued Green's functions~\cite{PhysRevB.104.165111}, it cannot be directly applied to bosonic systems in its original formalism. Quite recently, Nogaki \emph{et al.} suggest an ingenious trick to work around this limitation. Their basic idea is to introduce an auxiliary fermionic function~\cite{Nogaki2023bosonic}. Let us start with a bosonic Green's function $G(\tau)$ that satisfies the periodic condition $G(\tau+\beta) = G(\tau)$. One can construct an artificial anti-periodic fermionic Green's function $\tilde{G}(\tau)$ as follows:
\begin{equation}
\label{eq:auxiliary-fermionic-G-tau}
\tilde{G}(\tau) = \begin{cases}
G(\tau) &\quad (0<\tau<\beta) \\
-G(\tau + \beta) &\quad (-\beta<\tau<0)
\end{cases}
\end{equation}
Clearly, this auxiliary fermionic Green's function exhibits the same value as the bosonic Green's function in the range $0<\tau<\beta$. It is easy to prove that the relation between the bosonic spectral function $\rho(\omega)$ and the auxiliary fermionic spectral function $\tilde{\rho}(\omega)$ is as follows:
\begin{equation}
\label{eq:ABw-AFw-relation}
\rho(\omega) = \tilde{\rho}(\omega) \tanh(\beta\omega/2).
\end{equation}
Furthermore, the sum rule for $\tilde{\rho}(\omega)$ is given by:
\begin{equation}
\int_{-\infty}^{\infty} {\rm{d}} \omega~\tilde{\rho}(\omega) = G(\tau=0^+) + G(\tau=\beta-0^-).
\end{equation}
Given $\tilde{G}(\tau)$, it is easy to construct $\tilde{G}(i\nu_n)$ via direct Fourier transformation, where $\nu_n = (2n+1)\pi/\beta$ are the fermionic Matsubara frequencies. Since
\begin{equation}
\label{eq:gtilde}
\tilde{G}(i\nu_n) = \int_{-\infty}^{\infty}
    {\rm{d}} \omega \frac{\tilde{\rho}(\omega)}{i\nu_n - \omega},
\end{equation}
one can perform analytic continuation for $\tilde{G}(i\nu_n)$ via the standard NAC method to get $\tilde{\rho}(\omega)$. And then the bosonic spectral function $\rho(\omega)$ can be derived according to Eq.~(\ref{eq:ABw-AFw-relation}). This procedure has been outlined in Figure~1 in Ref.~\cite{Nogaki2023bosonic}.

\section{Benchmarks\label{sec:test}}

\subsection{Computational setups}

To benchmark the SPX and NAC methods, we consider four typical models, namely the Breit-Wigner model, the Gaussian mixture model, the resonance-continuum model, and the bottomonium model in the present investigation. The first three models provide analytic formulas for generating the exact spectral functions, denoted as $\rho(\omega)$. For the bottomonium model, we take the ``exact'' spectral function from Ref.~\cite{Aarts2014} as input. Using these spectral functions, one can create clean Euclidean data, denoted as $\mathcal{G}_{\text{clean}}$, using Eq.~(\ref{eq:kl})~\cite{nac_data}. To mimic the noise present in LQCD simulations~\cite{dipierro2001monte}, we manually add multiplicative Gaussian noises to the clean Euclidean data. The formula we use is as follows:
\begin{equation}
\label{eq:noise}
\mathcal{G}_{\text{noisy}} = \mathcal{G}_{\text{clean}} [1 + \delta N_{\mathbb{C}}(0,1)],
\end{equation}
where $\delta$ measures the noise level of the input data and $N_{\mathbb{C}}(0,1)$ represents complex-valued normal Gaussian noise~\cite{PhysRevB.107.075151}. In our subsequent analysis, unless explicitly stated, we set $\delta = 10^{-4}$ for the SPX method and $\delta = 0.0$ for the NAC method. We then supply the noisy Euclidean data, denoted as $\mathcal{G}_{\text{noisy}}$, into the SPX and NAC codes to extract the spectral functions. Finally, we compare the calculated spectral functions with the corresponding exact solutions.

The SPX method has been implemented within the \texttt{ACFlow} package~\cite{HUANG2023108863}. In this study, the number of poles ($N_p$) is fixed to 2000. For each test, we perform a total of $2 \times 10^3$ individual SPX runs. Each SPX run consists of $2 \times 10^5$ Monte Carlo sampling steps. The spectral functions generated in all SPX runs are gathered and the corresponding $\chi^2$ values are recorded. Assuming the mean value of the collected $\chi^2$ is denoted as $\langle \chi^2 \rangle$, we only retain the solutions whose $\chi^2$ values are smaller than $\langle \chi^2 \rangle / \alpha_{\text{good}}$, and apply them to calculate the averaged spectrum. Here, $\alpha_{\text{good}}$ is an adjustable parameter ($\alpha_{\text{good}} \ge 1.0$). Its optimal value is about 1.2.

Regarding the NAC method, we utilize another open-source toolkit, namely the \texttt{Nevanlinna.jl} package~\cite{nogaki2023nevanlinnajl}. In the present calculations, the lowest 100 Matsubara frequencies are kept as input. In order to avoid breaking the Pick criterion~\cite{Pick1917}, the optimal number of data points is automatically determined, which is denoted as $N_{\text{opt}}$, through a ``Pick Selection'' procedure in the algorithm. Usually a typical value for $N_{\text{opt}}$ is 10. During the simulated process, the Hardy basis function optimization algorithm is always enabled. The highest Hardy order, denoted as $H_{\text{max}}$, is set to be 50. To ensure numerical stability, the cutoff value of Hardy order, denoted as $H_{\text{cut}}$, should be determined automatically on a case-by-case basis. The $\lambda$ parameter, as seen in Eq.~(\ref{eq:loss-function}), is set to be $10^{-6}$.

In addition to the SPX and NAC methods, the classic MaxEnt method~\cite{ASAKAWA2001459} is also employed to yield analytic continuation results for comparison. We again utilize the \texttt{ACFlow} package, which provides a state-of-the-art implementation of the MaxEnt method~\cite{HUANG2023108863}. For the MaxEnt simulations, we usually choose a flat default model and calibrate the regularization parameter $\alpha$ by using the $\chi^2$-kink algorithm~\cite{PhysRevE.94.023303}. The starting and ending values of $\alpha$ are $10^{16}$ and $10^1$, respectively. The ratio between two successive $\alpha$ parameters, i.e. $\alpha_{i}/\alpha_{i+1}$, is 10. For the prior probability (the entropic term), we adopt the generalized Shannon-Jaynes entropy~\cite{JARRELL1996133,PhysRevB.44.6011}, but the Bayesian reconstruction entropy~\cite{PhysRevLett.111.182003,10.3389/fphy.2022.1028995} is also examined. The simulated results are in good agreement with each other. Thus, we only present results obtained with the generalized Shannon-Jaynes entropy in the following discussions.

\subsection{Breit-Wigner model\label{subsec:bw}}

\begin{figure*}[htbp]
\includegraphics[width=\textwidth]{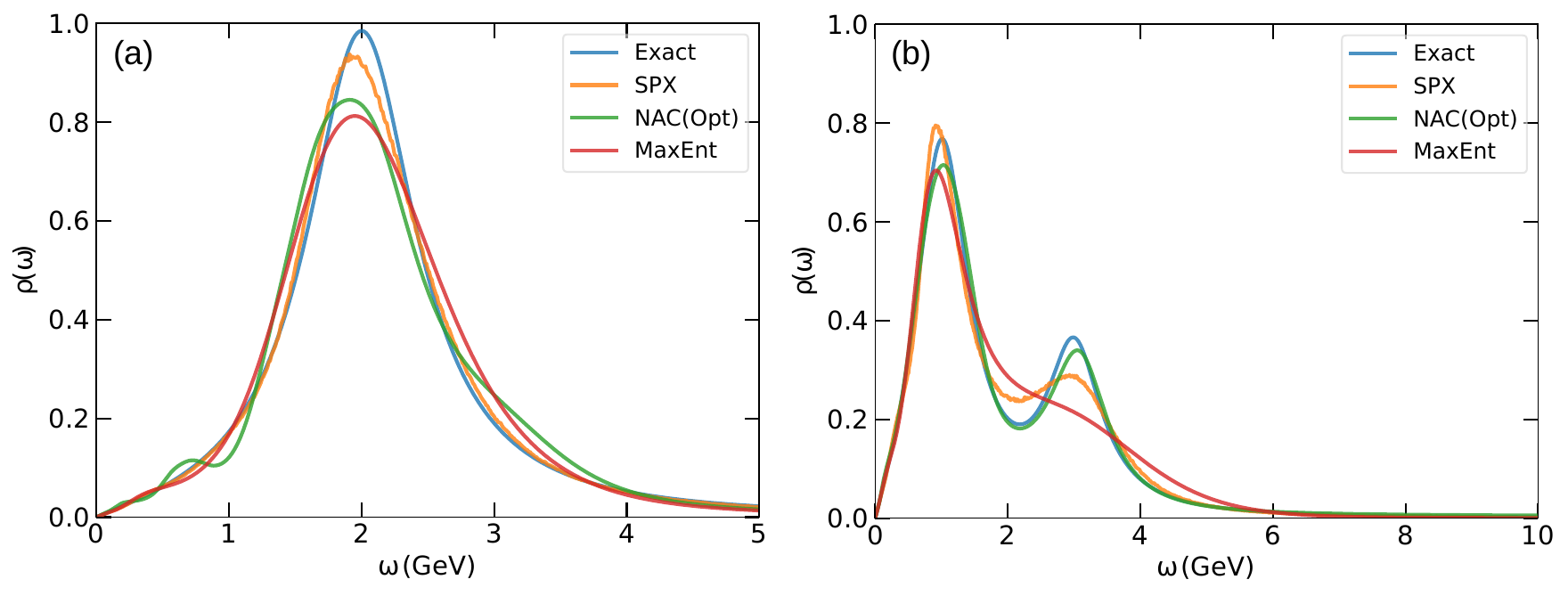}
\caption{Analytic continuations of the Breit-Wigner models. For the NAC method, the Hardy basis function optimization algorithm is adopted. (a) Single Breit-Wigner peak. $N_{\text{opt}} = 13$. (b) Two Breit-Wigner peaks. $N_{\text{opt}} = 14$. In panel (b) the spectra are scaled by a factor of 0.5 for a better view. \label{fig:Q01}}
\end{figure*}

The Breit-Wigner spectral function, obtained from a parameterization derived directly from one-loop perturbative quantum field theory~\cite{TRIPOLT2019129,PhysRevD.102.096001}, is expressed as follows:
\begin{equation}
\rho(\omega) = \frac{4A\Gamma \omega}{(M^2 + \Gamma^2 - \omega^2)^2 + 4 \Gamma^2 \omega^2},
\end{equation}
where $M$ represents the mass of the corresponding state, $\Gamma$ is the width, and $A$ is a positive constant. We start with a superposed collection of Breit-Wigner peaks. Specially, two typical scenarios are investigated in the present work: (1) Single Breit-Wigner peak (dubbed 1BW model) with $M = 2.0$~GeV, $\Gamma = 0.5$~GeV, and $A = 1.0$~GeV. (2) Two Breit-Wigner peaks (dubbed 2BW model) with $M_1 = 1.0$~GeV, $M_2 = 3.0$~GeV, $\Gamma_1 = \Gamma_2 = 0.5$~GeV, $A_1 = 0.8$~GeV, and $A_2 = 1.0$~GeV. The system temperature $T$ is fixed to be 0.02~GeV. The synthetic Euclidean data ($\mathcal{G}_{\text{noisy}}$) consists of 50 frequency points for both the SPX method and the MaxEnt method, and 100 frequency points for the NAC method. The analytical continuation results obtained by the SPX, NAC, and MaxEnt methods are illustrated in Figure~\ref{fig:Q01}.

Overall, the SPX method demonstrates better performance for the 1BW model. It captures precisely not only the height, but also the position of the Breit-Wigner peak. In comparison, the NAC and MaxEnt methods tend to overestimate the width and underestimate the height of the Breit-Wigner peak. Furthermore, the NAC method leads to an obvious oscillation phenomenon around 0.5~GeV. For the 2BW model, all three methods are able to reproduce the low-energy peak around 1.0~GeV, but encounter difficulty in resolving the high-energy peak near 3.0~GeV. Although the SPX method successfully identifies the position of the high-energy peak, its weight is not accurately resolved. The spectrum obtained by the NAC method is quite sensitive to the $\lambda$ parameter in this case. If $\lambda = 10^{-4}$ (it is the default choice of the code), the high-energy peak is shifted towards a higher energy ($\sim 4.5$~GeV) and broadened significantly. Only when $\lambda = 10^{-6}$, the high-energy peak is well reproduced. The spectrum obtained by the MaxEnt method is also not ideal. The high-energy peak is smeared out and replaced with a shoulder-like feature.

\subsection{Gaussian mixture model\label{subsec:gauss}}

\begin{figure}[t]
\includegraphics[width=\columnwidth]{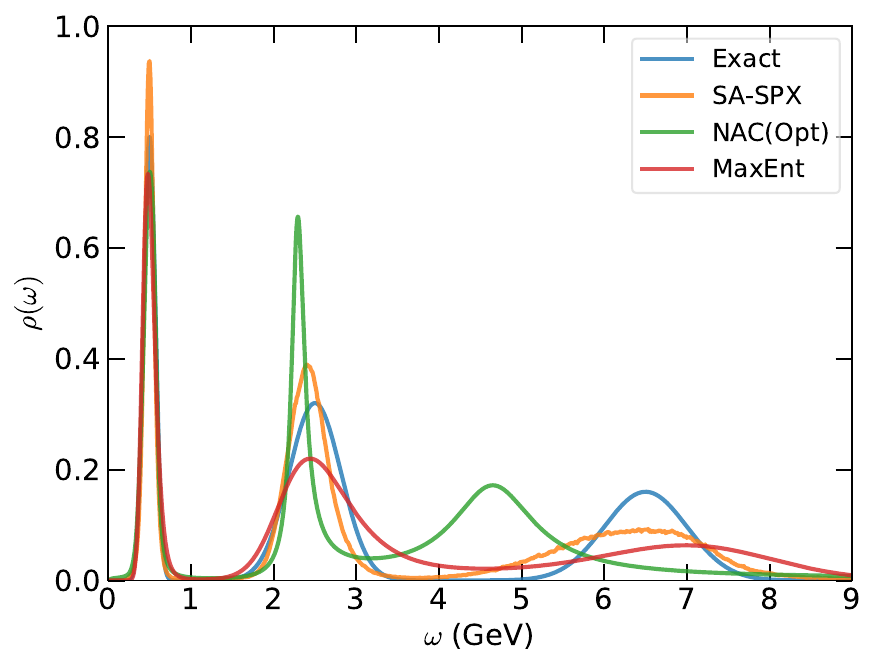}
\caption{Analytic continuations of the Gaussian mixture model. The spectra are scaled by a factor of 0.8 for a better view. As for the SPX method, the self-adaptive sampling algorithm is enable to obtain a more reasonable spectrum. Once the self-adaptive sampling algorithm is turned off, the obtained spectrum by the SPX method will resemble the one by the MaxEnt method. The spectrum should become smoother and almost featureless in the high energy region ($\omega > 4.0$~GeV). The NAC method is enhanced by the Hardy basis function optimization algorithm and $N_{\text{opt}} = 11$. Please see the main text for more details. \label{fig:Q06}}
\end{figure}

Just as its name implies, the spectral function of the Gaussian mixture model~\cite{ASAKAWA2001459} is a superposition of some Gaussian peaks. It can be expressed by the following equation:
\begin{equation}
\rho(\omega) = \sum_i A_i \exp\left[-\frac{(\omega - M_i)^2}{\Gamma_i}\right],
\end{equation}
where $A_i$, $M_i$, and $\Gamma_i$ represent the amplitude, position, and broadening of the $i$-th Gaussian peak. In this example, we consider a three Gaussian peaks model. The specific values for the model parameters are as follows: $A_1 = 1.0$ GeV, $A_2 = 0.4$ GeV, $A_3 = 0.2$ GeV, $M_1 = 0.5$ GeV, $M_2 = 2.5$ GeV, $M_3 = 6.5$ GeV, $\Gamma_1 = 0.01$ GeV, $\Gamma_2 = 0.2$ GeV, and $\Gamma_3 = 1.5$ GeV. The mock Euclidean data consist of 100 points, and $T = 0.02$ GeV. In the simulations, we have enhanced the SPX method by utilizing the self-adaptive sampling algorithm, which we refer to as SA-SPX~\cite{huang2023stochastic}. The results of the analytical continuation are presented in Figure~\ref{fig:Q06}.

It is expected that the true spectrum will exhibit three well-defined peaks. We find that all three methods are successful in recovering the low-energy sharp peak at $M_1$. However, resolving the two high-energy peaks presents some challenges. Specifically, for the SA-SPX method, it is able to roughly resolve the peak at $M_2$, but it tends to overestimate the width of the peak at $M_3$. In order to save the computational resources, the iteration number of the self-adaptive sampling algorithm is fixed to 10. Nevertheless, it should be emphasized that increasing the iteration number could further reduce the peak's width at $M_3$. In Section \ref{subsec:sasa}, we will delve into the combination of the self-adaptive sampling algorithm with the SPX method and discuss the usefulness of the SA-SPX method in resolving complex LQCD spectral functions. As for the NAC method, it produces a sharp peak around 2.1 GeV and a broad peak around 4.5 GeV, which are both in incorrect positions. More specifically, this method underestimates the energies of the two high-energy peaks. We also adjust the $\lambda$ parameter, but it doesn't help. Regarding the MaxEnt method, it can recover the peak at $M_2$, although with a larger width. However, it fails to resolve the peak at $M_3$, instead exhibiting a broad hump around $7.0 \pm 3.0$ GeV.

\subsection{Resonance-continuum model\label{subsec:rc}}

\begin{figure*}[t]
\includegraphics[width=\textwidth]{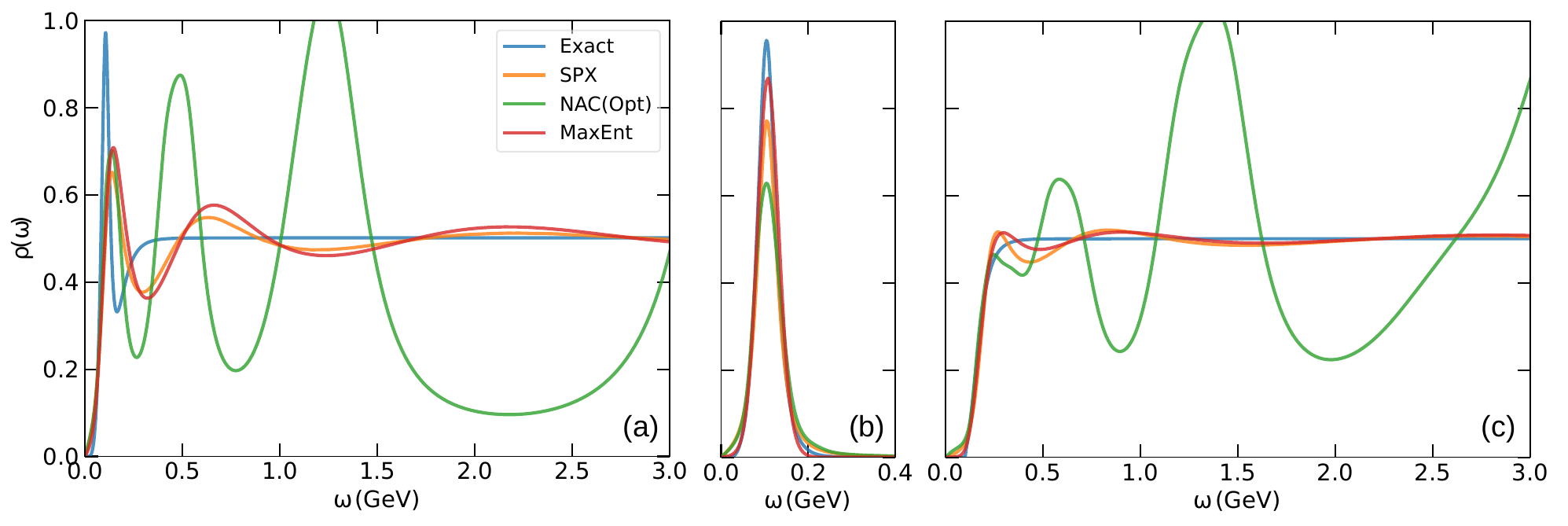}
\caption{Analytic continuations of the resonance-continuum models. The NAC method is enhanced with the Hardy basis function optimization algorithm. (a) Spectra of the resonance-continuum model. $N_{\text{opt}} = 12$. (b) Spectra of the resonance model. $N_{\text{opt}} = 10$. (c) Spectra of the continuum model. $N_{\text{opt}} = 13$. \label{fig:Q03}}
\end{figure*}

The resonance-continuum model is a physics-motivated model borrowed from References~\cite{PhysRevD.97.094503,PhysRevD.106.L051502}. The spectral function of the resonance-continuum model can be viewed as a nonlinear combination of the resonance part ($\rho_r$) and the continuum part ($\rho_c$):
\begin{equation}
\rho(\omega) = \xi_1(\omega) \rho_r(\omega) + \xi_2(\omega) \rho_c(\omega).
\end{equation}
Here, $\xi_1$ and $\xi_2$ are the mixing coefficients. Their definitions are as follows:
\begin{equation}
\xi_1(\omega) = \xi(\omega, M_r, \Gamma) \big[1 - \xi(\omega, M_r + \Gamma, \Gamma)\big],
\end{equation}
and
\begin{equation}
\xi_2(\omega) = \xi(\omega, M_c + \Gamma, \Gamma),
\end{equation}
where $\xi$ is a cutoff function:
\begin{equation}
\xi(\omega, M, \Delta) = \left[1 + \exp\left(\frac{M^2 - \omega^2}{\omega\Delta}\right)\right]^{-1}.
\end{equation}
It is used to smooth out the constructed spectral function. The resonance part of the spectral function is given by:
\begin{equation}
\rho_r(\omega) = C_r \omega^2 \left[\frac{\left(M_r^2 - \omega^2\right)^2}{M_r^2 \Gamma^2} + 1\right]^{-1},
\end{equation}
which follows a relativistic Breit-Wigner form. The continuum part of the spectral function is expressed as:
\begin{equation}
\rho_c(\omega) = \frac{3C_c}{8\pi\omega} \tanh\Bigg(\frac{\omega}{4T}\Bigg) \sqrt{\omega^2 - 4M_c^2}(2\omega^2 + 4M_c^2).
\end{equation}
In this example, the model parameters are $M_r = 0.10$ GeV, $M_c = 0.05$ GeV, $C_r = 2.0$ GeV, $C_c = 2.10$ GeV, and $\Gamma = 0.06$ GeV. The synthetic Euclidean data consist of 100 frequency points, and $T = 0.02$ GeV. In this study, we consider three different cases: (i) the resonance-continuum model, (ii) the resonance model, and (iii) the continuum model. The analytic continuation results are shown in Figure~\ref{fig:Q03}.

For the resonance-continuum model [see Fig.~\ref{fig:Q03}(a)], it is evident that the resonance peak at $M_r$ is approximately reproduced. However, in the continuum part ($\omega > 0.4$ GeV), both the SPX and MaxEnt methods exhibit moderate oscillations. These oscillations will decay when $\omega$ is increased. The NAC method results in huge oscillations, especially when $\omega$ is large. This unphysical feature can not be eliminated or suppressed by the Hardy basis function optimization algorithm. This fact suggests that the three analytic continuation methods do not accurately describe the continuum part. In the case of the resonance model [see Fig.~\ref{fig:Q03}(b)], all three methods successfully recover the location, width, and height of the resonance peak. As for the continuum model [see Fig.~\ref{fig:Q03}(c)], all three methods produce oscillating spectra. In particular, the NAC method leads to more pronounced oscillations. Even worse, these oscillations are enhanced with the increment of $\omega$. The only useful information that can be extracted from the calculated spectra is the location of the band edge.

\subsection{Bottomonium spectrum\label{subsec:bottom}}

\begin{figure}[t]
\includegraphics[width=\columnwidth]{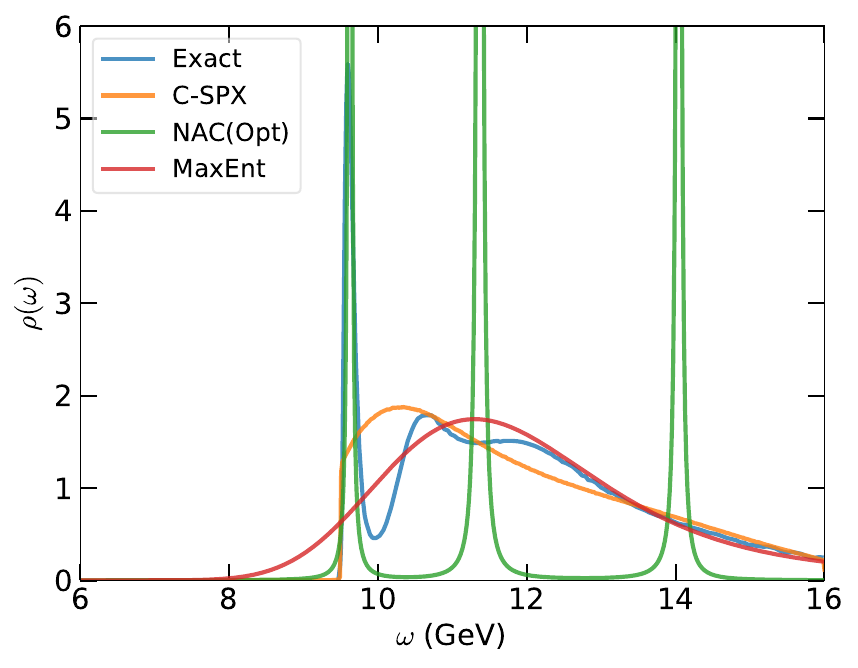}
\caption{Analytic continuations of the bottomonium correlation function. Here, the terminology ``C-SPX'' implies that the positions of the poles are restricted in the SPX simulations~\cite{c_spx,huang2023stochastic}. The Hardy basis function optimization algorithm is enabled for the NAC method and $N_{\text{opt}} = 8$. A shifted Gaussian function, instead of a constant, is used as the default model for the MaxEnt method~\cite{maxent}. See the main text for more details. \label{fig:Q07}}
\end{figure}

The ``exact'' bottomonium spectrum utilized in this study is taken directly from References~\cite{Aarts2014,ROTHKOPF20201}. It is generated by a $N_f = 2+ 1$ LQCD calculation. The temperature employed in the LQCD simulation is 201~MeV, which exceeds the deconfinement crossover temperature ($T_c$). The spectrum is specifically for the $\Upsilon$ channel. Initially, we synthesize the Euclidean data by Eq.~(\ref{eq:kl}) for the first 100 Matsubara frequencies. Then, random Gaussian noises are added by Eq.~(\ref{eq:noise}). In this example, we adopt the combination of the SPX method with the constrained sampling algorithm (dubbed C-SPX)~\cite{huang2023stochastic} to improve the performance. Specifically, locations for the randomly generated poles ($P_{\gamma}$) are restricted to the energy range: $\omega \in$ [9.5~GeV,16.0~GeV]~\cite{c_spx}. For the NAC method, the Hardy optimization trick is applied~\cite{PhysRevLett.126.056402}. Regarding the MaxEnt method, its default model is a shifted Gaussian function~\cite{maxent}. The analytic continuation solutions, together with the ``exact'' bottomonium spectrum, are illustrated in Fig.~\ref{fig:Q07}.

As is seen in Fig.~\ref{fig:Q07}, the ideal bottomonium spectrum consists of a single resonance peak at approximately 9.6 GeV and a ``rise-and-decay'' feature with two sizable bumps around 10.8 GeV and 12.0 GeV. By employing the constrained algorithm, the SPX method successfully resolves the left boundary of the resonance peak and captures the long tail of the ``rise-and-decay'' feature. However, it falls short in resolving the resonance peak and the two bumps. In Section \ref{subsec:sasa}, we will demonstrate the application of the self-adaptive sampling algorithm to cure this issue partly~\cite{huang2023stochastic}. For the NAC method, it accurately reproduces the resonance peak. But it fails to recover the desired ``rise-and-decay'' feature, which is instead replaced by two distinct peaks located at approximately 11.3 GeV and 14.0 GeV. According to our experience, though the spectrum obtained by the NAC method is not accurate enough, it can provide some hints about the energy range of the true spectrum (such as the position of the resonance peak in this case). We can use this information to refine subsequent simulations by imposing more reasonable constraints for the C-SPX method and more appropriate default models for the MaxEnt method. By employing the MaxEnt method, the resonance peak and the ``rise-and-decay'' pattern are smoothed out. Such that the calculated spectrum only features a Gaussian-like peak centered around 11.5~GeV. In addition to the shifted Gaussian model, we also benchmark alternative default models, such as the flat model, the shifted Lorentzian model, and the two Lorentzians model, etc. However, they do not contribute to improving the results.

\section{Discussions\label{sec:dis}}

\subsection{Robustness with respect to noisy data\label{subsec:noise}}

\begin{figure*}[ht]
\includegraphics[width=\textwidth]{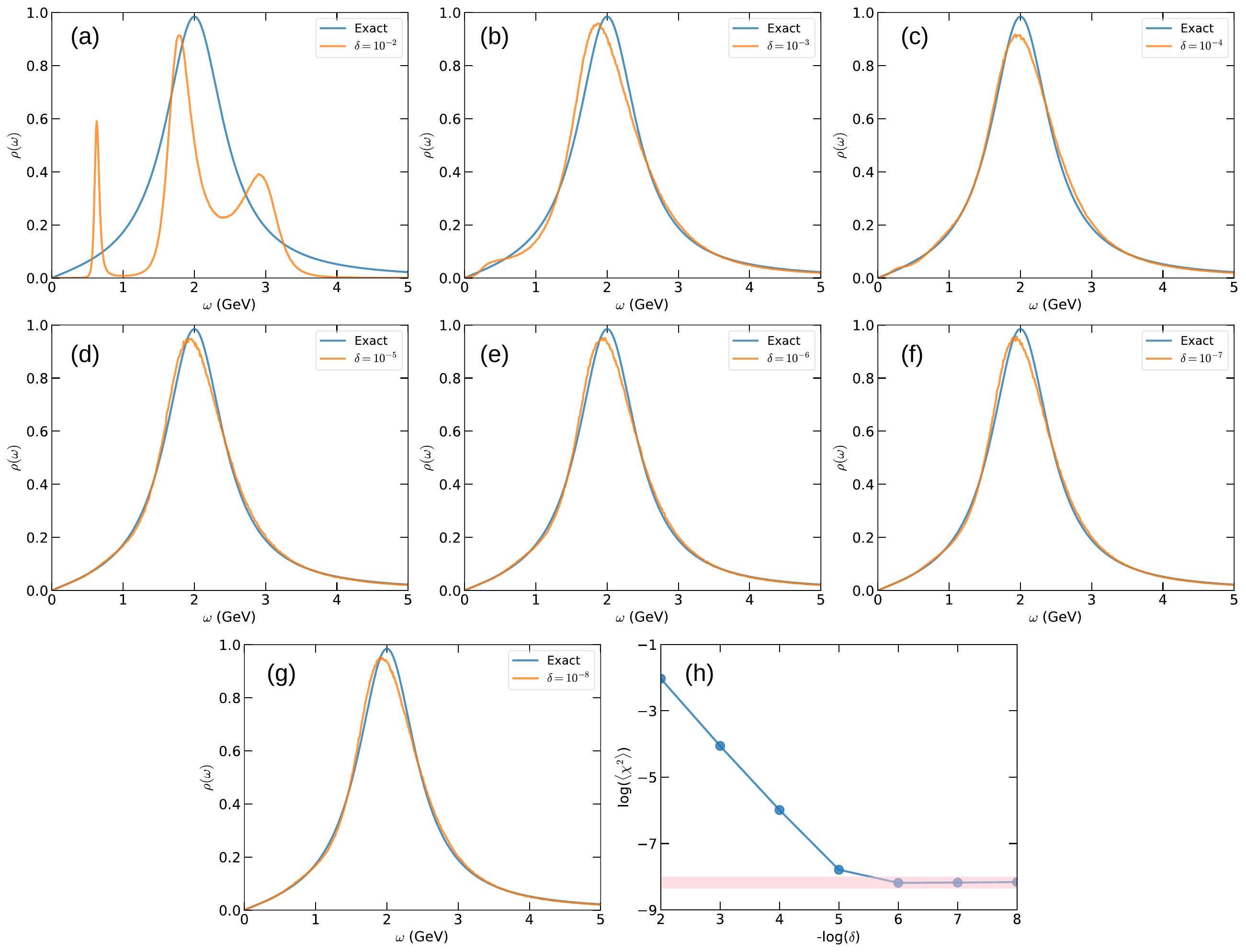}
\caption{Robustness of the SPX method with respect to the noisy LQCD data. Here we just consider the Breit-Wigner model (1BW model). The noise level $\delta$ is varied from $10^{-8}$ to $10^{-2}$. The other model parameters can be found in Sec.~\ref{subsec:bw}. (a)-(g) Dependence on noise level $\delta$ of calculated spectral functions. (h) The goodness-of-fit function $\chi^2$ as a function of the noise level $\delta$. The horizontal bar indicates the asymptotic value of $\log(\langle\chi^2\rangle)$. \label{fig:Q01N}}
\end{figure*}

\begin{figure*}[t]
\includegraphics[width=\textwidth]{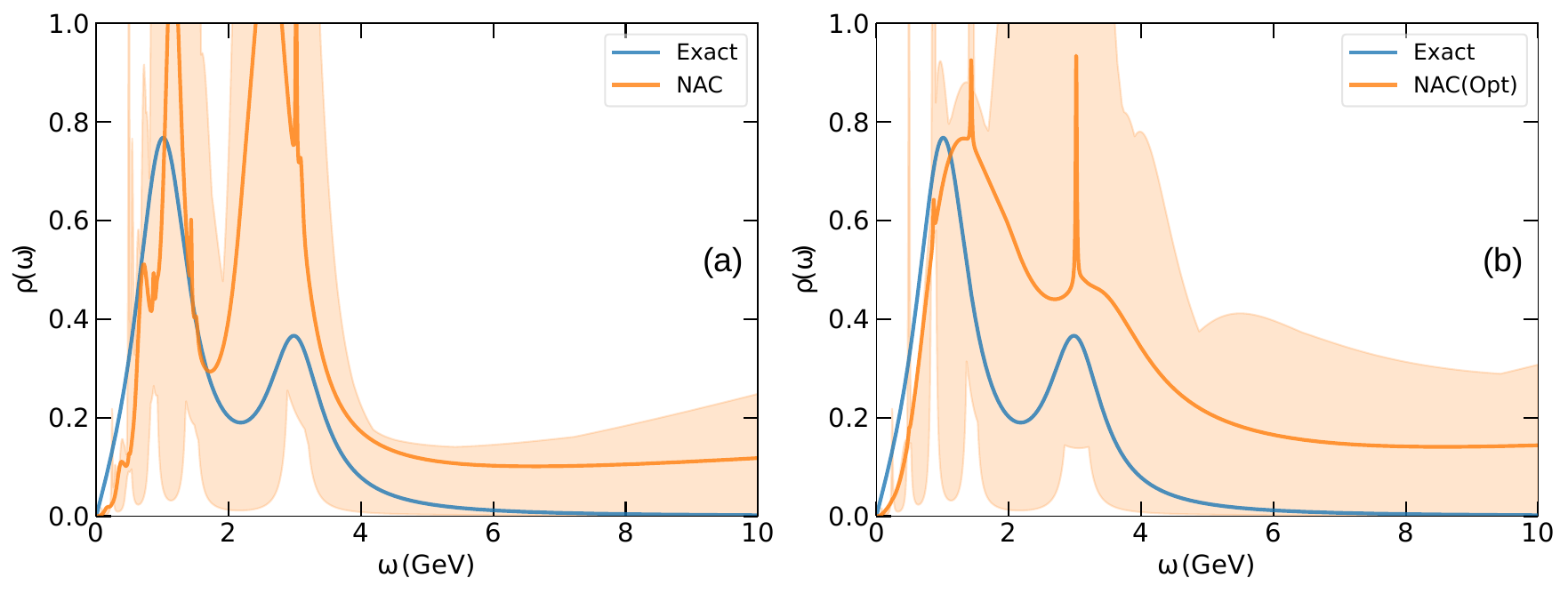}
\caption{Robustness of the NAC method with respect to the noisy LQCD data. Here we just consider the Breit-Wigner model (2BW model). The noise level $\delta = 10^{-8}$. The other model parameters can be found in Sec.~\ref{subsec:bw}. The darker shaded region denotes the window for the reconstructed spectral functions. The blue and yellow solid lines mean the exact spectrum and the averaged spectrum, respectively. (a) Without Hardy basis function optimization. (b) With Hardy basis function optimization. \label{fig:Q02N}}
\end{figure*}

In the previous work, the robustness of the SPX method in the presence of noisy data from quantum Monte Carlo simulations has been demonstrated~\cite{huang2023stochastic}. This study aims to reexamine the noise resilience of the SPX method when applied to noisy LQCD data. Let us take the 1BW model as an example (please refer to Section~\ref{subsec:bw} for the model parameters) to address this issue. The noise level is varied from $\delta = 10^{-8}$ to $\delta = 10^{-2}$. The analytic continuation results are displayed in Fig.~\ref{fig:Q01N}. Just as expected, the SPX method is highly robust to variations in noise levels of the input Euclidean data. For low noise levels ($10^{-4} \le \delta \le 10^{-8}$), the calculated spectra closely approximate the exact solution and manifest minimal deviation. At a moderate noise level ($\delta = 10^{-3}$), the calculated spectrum shows a light fluctuation around $\omega = 0.5$~GeV, and the main peak is shifted slightly (approximately 0.1~GeV) towards the lower energy region. For a high noise level ($\delta = 10^{-2}$), three sharp peaks emerge in the calculated spectrum. Apart from the peak at 1.8 GeV, the other peaks at 0.5 GeV and 3.0 GeV are unphysical. In Fig.~\ref{fig:Q01N}(h), a plot of $\log(\langle \chi^2\rangle)$ against $\log(\delta^{-1})$ is shown. Initially, $\log(\langle \chi^2\rangle)$ decreases linearly as $\log(\delta^{-1})$ increases from 2.0 to 5.0. Subsequently, it approaches to a constant value (approximately~-8.1) when $\log(\delta^{-1}) \ge 6.0$. This benchmark suggests that the SPX method remains robust when applied to noisy LQCD data, even with a moderately elevated noise level. Nonetheless, minimizing the noise level can enhance the performance of the SPX method.

Fei and Gull \emph{et al.} have pointed out that the NAC method requires high-precision input data to ensure the Pick criterion is not violated and the existence of the Nevanlinna interpolants~\cite{PhysRevLett.126.056402,PhysRevB.107.075151}. Thus, in the previous calculations, we just assume that the input Euclidean data are noiseless for the NAC method ($\delta = 0.0$). Now let us examine the noise resilience of the NAC method for synthetic LQCD data. For the sake of simplicity, the 2BW model is taken as a test-bed. The model parameters are presented in Section~\ref{subsec:bw}. The noise level $\delta$ is fixed to be $10^{-8}$. For each NAC run, the noisy part of the mock Euclidean data is always refreshed. We repeat the NAC simulations for 100 times with and without Hardy basis function optimization~\cite{Krantz1999}. Then we collect the calculated spectra and evaluate their arithmetic average. The analytic continuation results are shown in Fig.~\ref{fig:Q02N}. We confirm again that the NAC method is extremely sensitive to noise, irrespective of the Hardy basis function optimization algorithm. Small fluctuations in the input Euclidean data can lead to huge variations in the resulting spectra. Even though the noise level is quite small, the performance of the NAC method is not good. It fails to capture the major characteristics of the 2BW model, and produces some spurious peaks. If the noise level is further increased, its performance should deteriorate ulteriorly (not shown in Figure~\ref{fig:Q02N}).  

\subsection{Self-adaptive sampling algorithm for the SPX method\label{subsec:sasa}}

\begin{figure*}[t]
\includegraphics[width=\textwidth]{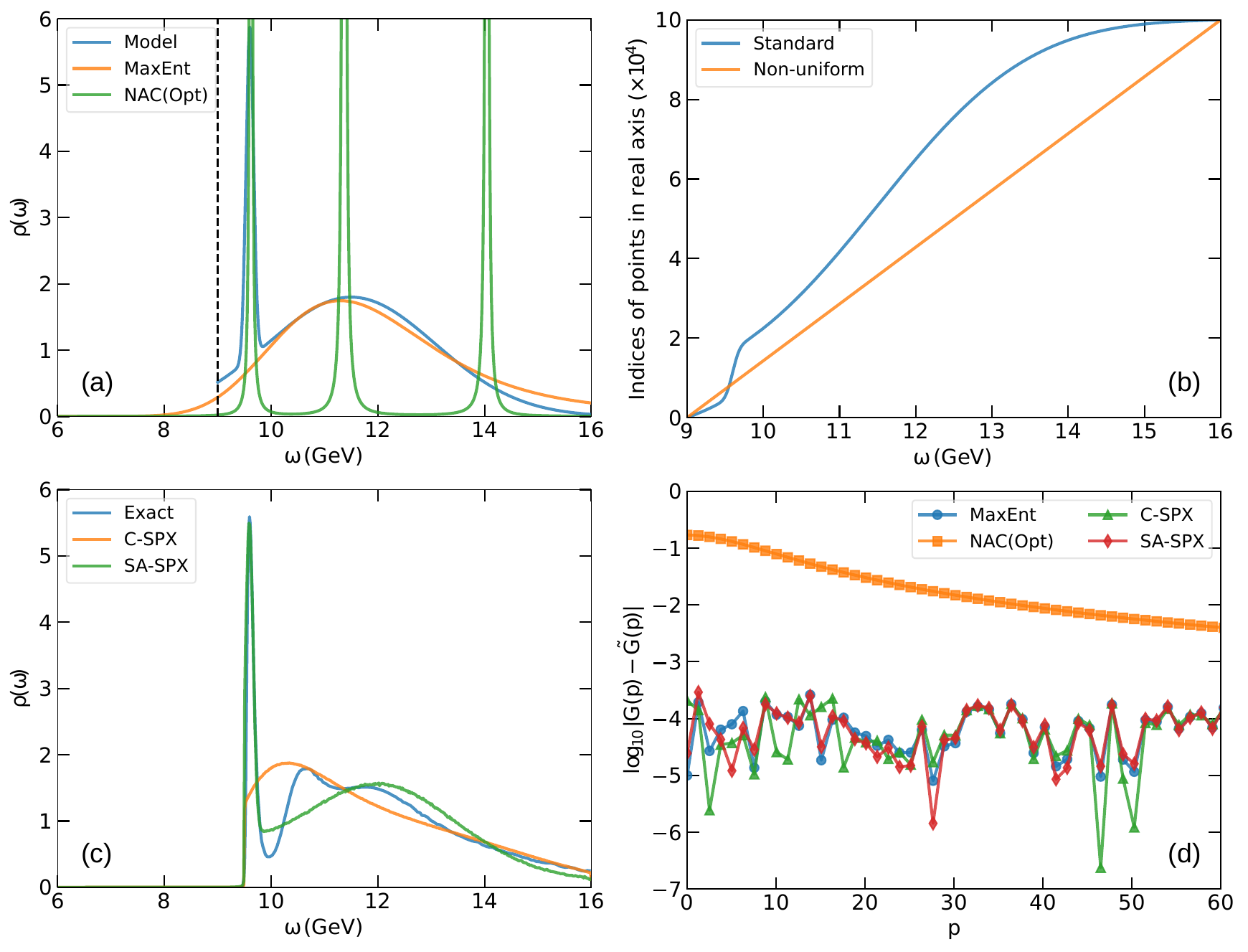}
\caption{Analytic continuations of the bottomonium correlation function. (a) Spectra obtained by the MaxEnt and optimized NAC methods. The information extracted from the two spectra is used to construct a reference model [the pseudo-spectrum $\rho_{\text{pseudo}}(\omega)$, see the solid blue line]. (b) Standard linear grid and non-uniform grid used in the C-SPX and SA-SPX simulations, respectively. Note that the non-uniform grid is constructed from the reference model via Eqs.~(\ref{eq:tos}) and (\ref{eq:new_mesh}). (c) Spectra obtained by the C-SPX and SA-SPX methods. (d) Error analysis for the reconstructed Euclidean data from the MaxEnt, optimized NAC, C-SPX, and SA-SPX methods. \label{fig:CSPX}}
\end{figure*}

In the SPX method, the poles should be placed in a very dense frequency grid. In general, such a frequency grid can be either uniform or non-uniform. So, some \emph{prior knowledge} about the spectrum and the physical system could be encapsulated in the form of the grid to improve the performance and usefulness of the SPX method. This has led to the development of the C-SPX method~\cite{huang2023stochastic}. Previous works have suggested that by modifying boundaries and grid interval distribution of the grid, the SPX method is capable of capturing complicated features in the spectra. However, it can be observed in Fig.~\ref{fig:Q07} that the C-SPX method, as well as the NAC and MaxEnt methods, fail to resolve the major characteristics of the bottomonium spectrum~\cite{Aarts2014,ROTHKOPF20201}. It implies that simple constraints on the spectral boundaries (or limitations on scopes of the poles) are not enough. We have to figure out a systematic way to refine the probability distribution of the poles to approximate the true spectrum. Next, we will demonstrate how to achieve this goal by a combination of the self-adaptive sampling algorithm and the SPX method (dubbed SA-SPX)~\cite{huang2023stochastic}.  

The main principle behind the SA-SPX method is to iteratively adjust the grid interval distribution. In consequence, the probability distribution of the poles is coordinated to approximate the true spectral density and the corresponding goodness-of-fit functional [see Eq.~(\ref{eq:loss})] is automatically minimized. This can be achieved by using the spectral density obtained from a previous SPX run or from other analytic continuation methods to update the grid. Now let us concentrate on the bottomonium model again~\cite{Aarts2014,ROTHKOPF20201}. Its parameters can be found in Section~\ref{subsec:bottom}. Initially, we generate the first frequency grid for the poles using the spectral functions obtained by the NAC and MaxEnt methods. From the calculated spectra, one can conclude that there is likely a sharp resonance peak around $\omega = 9.6$ GeV (with a band edge at approximately 9.5 GeV) and a broad feature ranging from 10.0 GeV to 16.0 GeV (see Fig.~\ref{fig:Q07}). Keeping these hints in mind, we try to design a pseudo-spectrum consisting of two Gaussian peaks:
\begin{equation}
\rho_{\text{pseudo}}(\omega) = \sum^2_{i = 1} A_i \exp\left[\frac{(\omega - M_i)^2}{\Gamma_i}\right],
\end{equation}
where $A_1 = 5.0$, $A_2 = 1.80$, $M_1 = 9.60$, $M_2 = 11.5$, $\Gamma_1 = 0.01$, and $\Gamma_2 = 5.0$. This pseudo-spectrum is illustrated in Fig.~\ref{fig:CSPX}(a). The first peak at $M_1$ originates from the resonance peak identified by the NAC method, while the second peak at $M_2$ is inspired by the spectrum obtained by the MaxEnt method. It is worth noting that these spectral parameters can be further adjusted to mimic more accurately the results obtained by the NAC and MaxEnt methods. This pseudo-spectrum serves as a reference model. To generate the frequency grid for the poles, we execute the following steps: (1) Calculate the integrated spectral function $\phi(\epsilon)$ via the equation:
\begin{equation}
\label{eq:tos}
\phi(\epsilon) = \int^{\epsilon}_{\omega_{\text{min}}} \rho_{\text{pseudo}}(\omega) d\omega,~\epsilon \in [\omega_{\text{min}},\omega_{\text{max}}].
\end{equation}
Here, $\omega_{\text{min}}$ and $\omega_{\text{max}}$ are the left and right boundaries of the spectrum, respectively. And $\epsilon$ represents a point within the interval $[\omega_{\text{min}},\omega_{\text{max}}]$. (2) Evaluate the new frequency grid $f_i$ by using the equation:
\begin{equation}
\label{eq:new_mesh}
f_i = \phi^{-1}(\lambda_i),~i = 1, \cdots, N_f,
\end{equation}
where $\lambda_i$ is a linear mesh in the interval $[\phi(\omega_{\text{min}}),\phi(\omega_{\text{max}})]$, and $N_f$ is the number of grid points. Now the boundaries for the grid are set as $\omega_{\text{min}} = 9.5$ GeV and $\omega_{\text{max}} = 16.0$ GeV. The resulting new grid, as displayed in Fig.~\ref{fig:CSPX}(b), is compared with the standard linear grid. Next, the newly generated grid is utilized to perform a SPX simulation from scratch. The calculated spectrum is then used to generate a newer frequency grid [at this time, the $\rho_{\text{pseudo}}(\omega)$ in Eq.~(\ref{eq:tos}) should be replaced with the calculated spectrum $\rho(\omega)$], and the SPX simulation is repeated. This iterative procedure is carried out until the obtained spectrum and frequency grid are converged. In our experience, 5 $\sim$ 10 iterations are typically sufficient for achieving convergence. Figure~\ref{fig:CSPX}(c) shows the results obtained by using the C-SPX and SA-SPX methods, as well as the exact spectrum. It is evident that the spectrum obtained with the SA-SPX method comes closer to the exact spectrum than that obtained with the C-SPX method. The sharp resonance peak, the small bump near 12.0~GeV, and the long tail of the rise-and-decay feature are well reproduced by using the SA-SPX method. The only missing characteristic is the bump near 10.8~GeV. Additionally, an error analysis about the reconstructed Euclidean data is presented in Fig.~\ref{fig:CSPX}(d). The goodness-of-fit function of the NAC method is the largest ($\chi^{2} \approx 0.01$ to $0.1$), while those of the MaxEnt, C-SPX, and SA-SPX methods are comparable ($\chi^2 \approx 10^{-6}$ to $10^{-4}$). This indicates that the spectrum obtained by the NAC method for this particular case is not reliable.

\subsection{Hardy basis function optimization for the NAC method\label{subsec:optnac}}

\begin{figure*}[t]
\includegraphics[width=\textwidth]{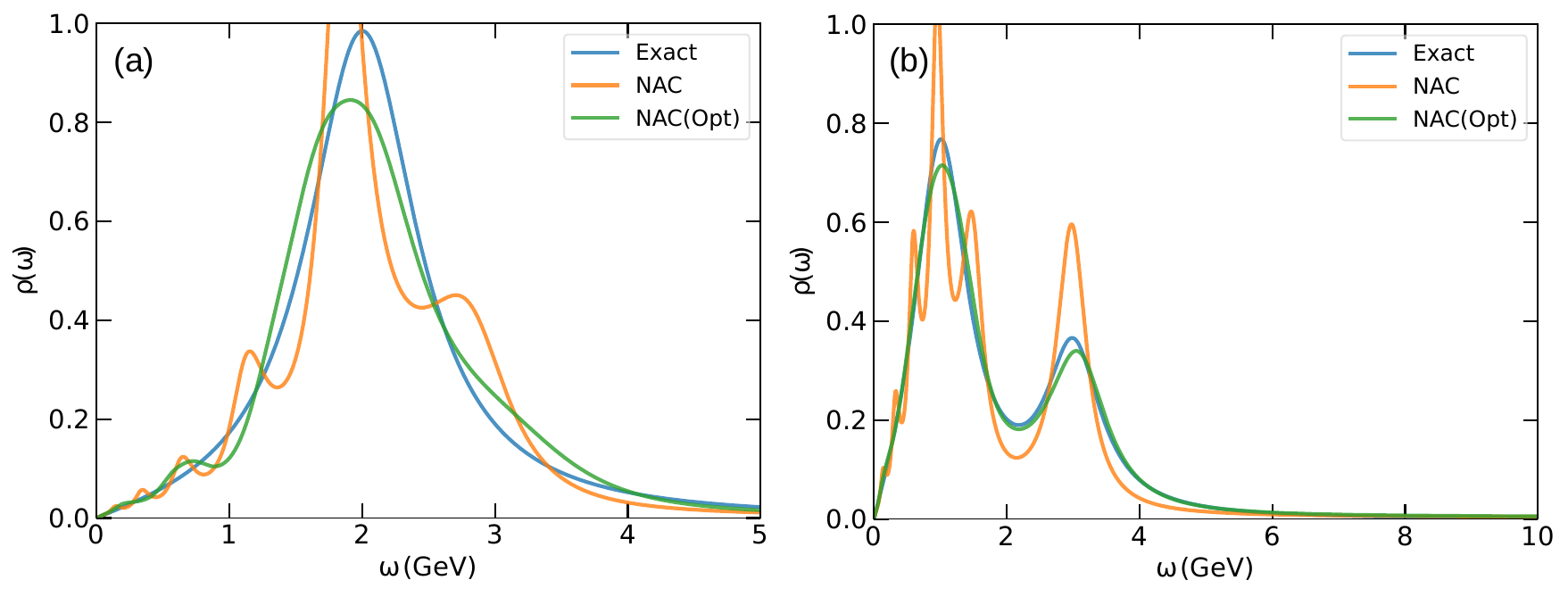}
\caption{A comparison for the NAC and optimized NAC methods. The Breit-Wigner model is used and the model parameters are the same as those presented in Sec.~\ref{subsec:bw}. The input Euclidean data are noiseless. (a) Results for the 1BW model. (b) Results for the 2BW model. \label{fig:NAC}}
\end{figure*}

As mentioned before, the iterative interpolation algorithm for the NAC method allows for the selection of an arbitrary contractive function, denoted as $\theta_{N_s+1}$, at the final step. In the literature, Fei and Gull \emph{et al.} proposed an optimized approach~\cite{PhysRevLett.126.056402}, in which $\theta_{N_s+1}$ is expanded using the Hardy basis and its conjugate generate functions~\cite{Krantz1999}, and the coefficients for the expansion are determined by minimizing a smoothness norm [see Eq.~(\ref{eq:loss-function})]. They demonstrated that using a constant value for $\theta_{N_s+1}$ is apt to yield spectral functions with oscillations, whereas the optimized algorithm is useful for eliminating these oscillations and generating smoother spectral functions. Until now we just employ the optimized NAC approach for analytic continuations. However, we wonder whether the Hardy basis optimization is always better than the standard option. In order to answer this question, we perform additional tests for the Breit-Wigner model by using the NAC method. We compare the analytic continuation results obtained with a constant $\theta_{N_s+1}$ and the optimized $\theta_{N_s+1}$ (see Fig.~\ref{fig:NAC}). As anticipated, the optimized $\theta_{N_s+1}$ suppresses oscillations in a large measure and yields smoother spectra, especially for the 2BW model. However, we also notice that the performance of the Hardy basis function optimization algorithm strongly depends on the value of the $\lambda$ parameter. The optimized NAC method tends to make a wrong estimation about the location of the high-energy peak if the $\lambda$ parameter is not reasonable. Therefore, if we know nothing about the basic features of the spectra, perhaps a constant $\theta_{N_s+1}$ is a much safer choice. 

\section{conclusion\label{sec:con}}

In the present work, we conduct a systematic investigation of two newly developed methods, namely the SPX method and the NAC method, for analytically continuing for the mock LQCD data. We treat four exact spectral functions, which are derived from physically motivated models or realistic LQCD simulations, including the Breit-Wigner model, the Gaussian mixture model, the resonance-continuum model, and the bottomonium model. We use the exact spectral functions to build clean Euclidean data by numerical integration. And later the statistical noises are added. The synthetic Euclidean data are used as input and then transformed back to the real axis using different analytic continuation methods. By comparing the results with the exact spectra, we are able to assess the accuracy of these methods.

The SPX method is generally capable of resolving the major features of the spectral functions involved in this study. However, it encounters difficulties when dealing with spectra that exhibit a wide platform, such as the continuum model (see Sec.~\ref{subsec:rc}), or when two features are too close together, such as the bottomonium spectrum (see Sec.~\ref{subsec:sasa}). We believe that these difficulties can be partially overcome with the help of the constrained sampling algorithm and the self-adaptive sampling algorithm. The SPX method demonstrates good noise tolerance and exhibits robustness with respect to moderate noise levels. Overall, the performance of the SPX method is comparable to that of the commonly used MaxEnt method. In cases where the spectral function is complicated, the SPX method could outperform the MaxEnt method due to its ability to incorporate prior information about the spectrum into the frequency grid for the poles. This grid could be iteratively refined to obtain better spectrum. 

As for the NAC method, it is found to be numerically unstable even for input Euclidean data with extremely low noise level ($\delta = 10^{-8}$). This drawback greatly limits the application of the NAC method in the LQCD simulations. Additionally, we observe that the Hardy basis optimization for $\theta_{N_s+1}$ sometimes produces worse results when compared to those obtained with constant $\theta_{N_s+1}$. Although the Hardy basis optimization can suppress possible oscillations in the spectrum, it tends to yield incorrect estimations for the positions of the high-energy peaks if the $\lambda$ parameter is not optimal. Therefore, better basis functions for expanding $\theta_{N_s+1}$ are highly desirable. Or else we need a smart algorithm to determine the optimal $\lambda$. Nonetheless, the NAC method still proves its usefulness in analytic continuations of LQCD simulation data, as it allows for quick yet accurate estimations of the positions of the low-energy band edges and the resonance peaks. These important clues can then be used to construct a reference model for the probability distribution of the poles, which is subsequently utilized by the constrained SPX method.

\begin{acknowledgments}
The authors thank Prof. Lei Wang for fruitful discussions. This work is supported by the Innovation Foundation of China Academy of Engineering Physics (No.~CX20200033), the National Natural Science Foundation of China (No.~12274380 and No.~11934020), the National Key Projects for Research and Development of China (No.~2021YFA1400400), the China Postdoctoral Science Foundation (No.~2021TQ0355), and the Special Research Assistant Program of Chinese Academy of Sciences.
\end{acknowledgments}

\bibliography{spxd}

\begin{thebibliography}{70}
\bibitem [{nac()}]{nac_data}%
  \BibitemOpen
\bibfield  {number} {  }\href@noop {} {}\bibinfo {note} {To prepare input data
  for the NAC method, one at first converts $\rho(\omega)$ to
  $\tilde{\rho}(\omega)$ via Eq.~(\ref{eq:ABw-AFw-relation}), and then
  calculates $\tilde{G}(i\nu_n)$ via Eq.~(\ref{eq:gtilde})}\BibitemShut
  {NoStop}%
\bibitem [{c_s()}]{c_spx}%
  \BibitemOpen
  \href@noop {} {}\bibinfo {note} {The left and right boundaries for the C-SPX
  method can be deduced from the spectrum obtained by the NAC method. Actually,
  we perform analytic continuation by the NAC method at first. From the
  obtained spectrum, we can make a rough estimation about the boundaries of the
  true spectrum. Then this information can be utilized to customize the
  constraints for the C-SPX method. Please see Ref.~\cite{huang2023stochastic}
  for more details.}\BibitemShut {Stop}%
\bibitem [{max()}]{maxent}%
  \BibitemOpen
  \href@noop {} {}\bibinfo {note} {The center of the Gaussian peak is fixed to
  be 11.5~GeV, which is learnt from the spectrum obtained by the NAC method.
  Please refer to Ref.~\cite{c_spx} as well.}\BibitemShut {Stop}%
\end{thebibliography}

\end{document}